\documentclass[a4paper,fleqn,usenatbib]{mnras}
\usepackage{amsmath,amssymb}

\usepackage{upgreek}
\usepackage{txfonts}
\usepackage[T1]{fontenc}
\usepackage{ae,aecompl}

\usepackage{graphicx}	% Including figure files
\usepackage{amsmath}	% Advanced maths commands
\usepackage{amssymb}	% Extra maths symbols

\title[Giant molecular clump and molecular cloud collisions]{Dynamical cooling 
of galactic discs by molecular cloud collisions -- Origin of giant clumps in
gas-rich galaxy discs}

\author[Guang-Xing Li]{
Guang-Xing Li$^{1}$\thanks{E-mail: gxli@usm.lmu.de (USM)}
\\
$^{1}$University Observatory Munich, Scheinerstrasse 1, D-81679 M\"unchen,
Germany\\}
\date{Accepted XXX. Received YYY; in original form ZZZ}

\pubyear{2015}

\begin{document}
\label{firstpage}
\pagerange{\pageref{firstpage}--\pageref{lastpage}}
\maketitle

% Abstract of the paper
\begin{abstract}
% Astrophysical fluid systems are characterized by high Reynolds numbers,
%  yet gravity is important in many situations.
Different from Milky-Way-like galaxies,
discs of gas-rich galaxies are clumpy.
It is believed that the
clumps form because of gravitational instability. However, a necessary
condition for gravitational instability to develop is that the disc must
dissipate its kinetic energy effectively, this energy dissipation (also
called cooling) is not well-understood. We propose that collisions
(coagulation) between molecular clouds dissipate the kinetic energy of the discs, which leads to a dynamical cooling.
The effectiveness of this dynamical cooling is quantified by the dissipation
parameter $D$, which is the ratio between the free-fall time $t_{\rm ff}\approx 1/ \sqrt{G
\rho_{\rm disc}}$ and the cooling
time determined by the cloud collision process $t_{\rm cool}$. This ratio is
related to the ratio between the mean surface density of the disc $\Sigma_{\rm 
disc}$ and the mean surface density of molecular clouds in the disc $\Sigma_{\rm
cloud}$. When $D <1/3$ (which roughly
corresponds to $\Sigma_{\rm disc} < 1/3 \Sigma_{\rm cloud}$), cloud collision
cooling is inefficient, and fragmentation is suppressed. When $D > 1/3$
(which roughly corresponds to $\Sigma_{\rm disc} > 1/3 \Sigma_{\rm cloud}$), cloud-cloud collisions lead to a
rapid cooling through which clumps form.  On smaller scales, cloud-cloud
collisions can drive molecular cloud turbulence.
This dynamical cooling process can be taken into account in numerical
simulations as a subgrid model to simulate the global evolution of disc
galaxies.

\end{abstract}

% Select between one and six entries from the list of approved keywords.
% Don't make up new ones.

%%%%%%%%%%%%%%%%%%%%%%%%%%%%%%%%%%%%%%%%%%%%%%%%%%

\section{Introduction}
Star formation occurs in the cold phase of the galactic interstellar
medium (ISM), which exhibits diverse morphologies in different galaxies. The
Milky Way is an example where molecular gas is mostly distributed in molecular
clouds whose sizes are much smaller than the disc scaleheight.
In contrast to this, in high-redshift, gas-rich galaxies, molecular gas seems to
concentrate in giant (kpc-size), massive clumps, and majority of the star
formation occurs in the clumps 
\citep{1996MNRAS.279L..47A,1998Natur.392..253N,2005ApJ...627..632E,2008ApJ...687...59G,2015ApJ...800...39G,2009ApJ...701..306E,2009ApJ...706.1364F}
\footnote{Sometimes the giant clumps appear in chains, and these galaxies are called ``chain galaxies''
\citep{1995AJ....110.1576C,1996AJ....112..359V,2004ApJ...604L..21E,2004ApJ...603...74E,2005ApJ...620..564C},
and \citet{1996ApJ...465L...9D} suggested that the chain-like morphology is due the projected view of an edge-on galaxy.}.
{ Unlike molecular clouds, the clumps have sizes that are often
comparable to the disc scaleheight. How do they form?}

One promising clump formation path is the gravitational instability,
where the growth of perturbations on the disc scale leads to structure
formation.
In fact, the sizes of the observed clumps are often comparable to the Toomre
length \citep{2017arXiv170300458F}, suggesting that gravitational instability
is at work. The short-time evolution of the gravitational instability is determined by the Toomre $Q$ parameter, where $Q_{\rm disc}= \sigma_{\rm v, disc}\Omega_{\rm disc}/ \uppi G \Sigma_{\rm disc}$,  $\sigma_{\rm v, disc}$  is the velocity dispersion, $\Sigma_{\rm disc}$ is the disc surface density, and $\Omega_{\rm disc}$ is disc angular frequency. In Toomre's formalism, fragmentation is
driven by gravity (which is characterized by $\Sigma_{\rm disc}$),
which is balanced by a combination of supported caused by local motions
$\sigma_{\rm v, disc}$ and shear
($\Omega_{\rm disc}$, which is roughly the epicyclic frequency).
Discs with $Q_{\rm disc}<1$ are unstable against non-axisymmetric instabilities
and should fragment.

It has been believed that the clump formation is  largely controlled by the Toomre parameter
\citep{2005MNRAS.363....2K,2009Natur.457..451D,2009ApJ...703..785D,2016MNRAS.456.2052I}
or a Toomre-like formalism, e.g. \citet{2010MNRAS.407.1223R}. 
However, the Toomre formalism seems to provide a still incomplete
description to the clump formation process:
observationally, the connection  between a smaller Toomre $Q$ parameter and a clumpy disc has not
been established. On the contrary, observations indicate that
galactic discs tend to have  $Q_{\rm disc}\approx 1$ with an observed
uncertainty of a factor of two, and this is almost independent on whether a disc
is clumpy or not \footnote{There
are some variations of the Toomre $Q$ parameter in different galaxies, with
clumpy galaxies having smaller $Q$  \citep[e.g.][]{2014ApJ...790L..30F}. 
But it remains a theoretical issue how to define the Toomre $Q$ in a disc that
consists of both gas and stars accurately
\citep{1992MNRAS.256..307R,1994ApJ...427..759W,2011ApJ...737...10E,2011MNRAS.416.1191R,2016MNRAS.456.2052I}.
Therefore, some of the observed variations in Toomre $Q$ might arise from this
theoretical uncertainty.
Besides, the variation is
much smaller than the difference in the $D$ parameter that what we discuss in
this paper, and we believe that it is the difference in the $D$ parameter that
leads to different evolution.}
\citep{2009A&A...495..795H,2010MNRAS.406..535P,2012MNRAS.421..818C,2014ApJ...790L..30F,2013MNRAS.433.1389R,2017arXiv170102138R}
\footnote{Where, to properly evaluate the
$Q$ parameter, one should properly combine the dynamical contribution of the
stars and the gas, see e.g. \citet{2013MNRAS.433.1389R,2017arXiv170102138R}.}.
This cannot be explained within the Toomre formalism, but is believed to the
the result of a self-regulated disc evolution. Since the discs are already self-regulating
systems, a better theory of clump formation should take this 
into account.

 Apart from this, more recent theoretical studies also  suggested
 that the efficiency of disc cooling plays a more fundamental role in clump
 formation.
 \citet{2001ApJ...553..174G,2003ApJ...597..131J} carried out a
 set of local, shearing box simulations of disc evolution, and found that (a) the balance
 between heating and cooling processes tends to regulate the disc to a state
 where $Q_{\rm disc}\approx 1 $ .
 They also found that (b) it is the cooling time that determines the long-time evolution of such discs:
 if cooling is inefficient
(cooling time is long compared to the dynamical time), the disc should evolve
into a gravo-turbulent state where
 turbulence maintained by accretion can effectively suppress the overly-rapid
 fragmentation, and the disc is dominated by fluffy, filamentary structures that
 are constantly smeared apart by rotation; when
cooling is efficient (cooling time is short compared to the dynamical time), gas
in the disc rapidly dissipates kinetic energy and fragments into a few giant
clumps.

 The theory of \citet{2001ApJ...553..174G} seems to provide a good description
 the evolution of disc galaxies. First,
both simulations
\citep[e.g.][]{2012MNRAS.421.3488H,2013A&A...555A..72L,2016ApJ...827...28G} and
observations
\citep{2009A&A...495..795H,2010MNRAS.406..535P,2012MNRAS.421..818C,2014ApJ...790L..30F}
seem to indicate that discs tend to have $Q_{\rm disc}\approx 1$ 
, consistent with self-regulation.
Secondly, we do observe a similar dichotomy in the morphology of the observed
galaxies: in  \citet{2001ApJ...553..174G}, a disc can be either filamentary or clumpy, and this seems to correspond to the
 fact that in galaxies gas either organises into smaller clouds that can be
 aligned on a larger scale (e.g. the
 Milky Way where they form larger filaments) or into giant clumps (e.g. the
 gas-rich clumpy galaxies).
 This resemblance poses the question: can we explain the morphological
 difference of galaxies using the theory of \citet{2001ApJ...553..174G}?
What is the major cooling mechanism in these discs?

The cooling of a galactic disc remains an open issue. We emphasize that the disc
is dynamically supported by some motions characterized by a velocity dispersion
$\sigma_{\rm v, disc}$, and the disc-scale cooling should be able to reduce
this velocity dispersion.
When the disc is supported by thermal pressure, $\sigma_{\rm v, disc}$ is related to the temperature of the gas; when the disc
is supported by some random motions, $\sigma_{\rm v, disc}$ is the velocity
dispersion of these random motions. Previously, the dominate cooling mechanism
of these discs is not well-identified:
 \citet{2009ApJ...703..785D,2016ApJ...830L..13M} identified the
radiative cooling time of the atomic ISM as the cooling time relevant for disc fragmentation, and
\citet{2002ApJ...577..197W,2010ApJ...712..294E,2010A&A...520A..17K,2011ApJ...737...10E,2014MNRAS.438.1552F}
\footnote{They are based on earlier models, such as
\citet{1974MNRAS.168..603L,1987MNRAS.225..607L,1987ApJ...320L..87L}.}
assumed galactic disc turbulence model, and identified the turbulence
dissipation time as the disc cooling time. We argue that
neither are adequate. This is because in a real galactic disc (we will be
focusing on gas-rich discs where more than half of the gas is molecular, see
e.g.
\citet{2010Natur.463..781T}), the local dynamical support comes from the quasi-random motions of the molecular gas. To cool down a disc one must reduce this random motion.
\citet{2009ApJ...703..785D} mistakenly believed that relevant cooling
time is the radiative cooling time of the atomic gas. However, they
overlooked the critical fact that even if one has successfully converted the
atomic gas into the molecular gas through the radiative cooling, the disc can still be supported
by motions of the molecular ISM. The
kinetic energy contained in these motions cannot be reduced with the radiative
cooling, and additional dynamical cooling channels are needed. Another set of
models are based on the picture of galactic disc turbulence \citep{2002ApJ...577..197W,2010ApJ...712..294E,2010A&A...520A..17K,2011ApJ...737...10E,2014MNRAS.438.1552F}.
These models correctly identified that to cool down the disc, one must
reduce the velocity dispersion of the ISM. However, they made the assumption
that the dynamics of the multi-phase ISM can be described by the turbulence
model -- an assumption that is widely accepted yet not justified: the standard
theory of turbulence has only been shown to be valid for non-self-gravitating gas  with well-defined equation of states.
 In a real galactic disc,
gas is separated into different phases where molecular gas has a 
density
that is two orders of magnitudes larger than the density of the ambient ISM
(See also Sec. \ref{sec:general}).
This density contrast makes many assumptions of the galactic disc turbulence
model invalid. For example, in the standard picture of galactic disc turbulence,
ideally, one requires energy to cascade from the larger scales continuously down
to the Kolmogorov microscales where it dissipates into heat. But in a real
galactic disc, due to the density contrast between the molecular clouds and
ambient gas as well as the fact that molecular gas forms centrally-condensed structures, it is
practically very difficult for the kinetic energy in the warm ISM to cascade
into molecular clouds \footnote{E.g. 
\citet{2016ApJ...824...41I} found that the due to the density contrast between
the molecular ISM and the warm neutral medium, it is difficult for supernova
feedback to drive molecular cloud turbulence; in the analytical study of
\citet{2017MNRAS.465..667L}, the author found that when a region which has
$\rho\sim r^{-2}$ where $\rho$ is the density and $r$ is the radius, it is
difficult for turbulence to cascade from the outside to the inside. }. Another consequence of such a density contrast is that if a dense
molecular cloud travels in a galactic disc with a diffuse ambient medium, in most cases the cloud would neglect the drag  from the ambient medium and travel on its own \citep[as has
been estimated by][]{1983Ap&SS..89..177M}. One must take this dynamical
detachment induced by phase separation into account in realistic galactic disc
models. Previously, some models simply stepped over the cooling issue by
assuming isothermal equations of states \citep{2015MNRAS.448.1007B,2016ApJ...819L...2B}.
They simulated gravitationally unstable isothermal discs, and found that the
disc first fragments into rings and then into clouds,
and in the final step the clouds merge to form clumps. The possibility of
merging clouds found in their simulations is an interesting one. However, since
they have assumed an isothermal equation of state, how cooling leads to disc fragmentation remains
unclear.

We consider a picture where a galactic disc
consists of an ensemble of colliding molecular clouds.{
 We assume the existence of these clouds, and study the consequence of
cloud-cloud collisions on the evolution of the galactic discs.} We argue that a
dynamical cooling occurs because of the collisions between the clouds, and on the macroscopic level this cooling
can lead to the formation of the clumps.
The previous studies on cloud-cloud collisions in galaxies
\citep{1983ApJ...271..604K,1983Ap&SS..89..177M,1986PASJ...38...95T,2000ApJ...536..173T,2002A&A...382..872V,2008MNRAS.391..844D}
mainly considered the evolution of clouds in fixed galactic potentials and focused on properties of the clouds.
{ In their restricted settings, clouds are usually not allowed to  back-react
on the disc dynamics.}
These models are valid for Milky Way like galaxies where disc surface
densities are low, but are not valid for gas-rich discs.
In this paper, we carry their analyses into the non-linear regime, where the
cloud-cloud collisions that occur on smaller scales
are able to back-react on the disc dynamics on the large scale. When the
collision rate is high, clumps form as a result of the collective motion of the
molecular clouds included by cloud collision cooling.

\begin{table*}
\begin{tabular}{l | l }
  \hline			
Variable Name & Definition  \\
  \hline			
 \multicolumn{2}{|c|}{Microscopic variables}\\
 \hline
 $r_{\rm cloud}$ & Typical  size of a molecular cloud \\
 $m_{\rm cloud}$ & Typical mass of a molecular cloud \\
 $\Sigma_{\rm cloud}$ & Typical surface density of a molecular cloud. 
 $\Sigma_{\rm cloud}\approx  \Sigma_{\rm cloud, crit} $  \\
  $\Sigma_{\rm cloud, crit}$ & Critical surface density of a molecular cloud to
  survive against photodissociation (Eq.
  \ref{eq:sigma:cloud})
  \\
 $\sigma_{\rm v, cloud}$ & Velocity dispersion of a molecular cloud \\
 \hline
  \multicolumn{2}{|c|}{Macroscopic variables}\\
 \hline
 $n_{\rm cloud}$ & Mean number density of molecular clouds\\

 $\sigma_{\rm v,  disc}$ & Velocity dispersion of the galactic disc \\

 $\rho_{\rm disc}$ &  Mean density of the galactic disc \\
 $\rho_{\rm disc, crit}$ & Critical density for
 effective cloud collision cooling\\
 $\Sigma_{\rm  disc}$ & Surface density of the galactic disc \\
 $t_{\rm cool} \approx t_{\rm collide}$ & Cloud collision cooling time 
 $\approx$ cooling time \\
 $\lambda_{\rm cloud}$ & Mean free path of molecular clouds \\
 $G$ & Gravitational constant \\
 $Z$  & Metallicity \\
 $G_0$ & Interstellar radiation field\\ 
 $\Omega_{\rm kep}$ & Angular frequency of galactic  (which is approximately
 the epicyclic frequency)
 \\
 $t_{\rm kep}$ & Disc dynamical time $t_{\rm kep}\approx 3 \Omega_{\rm
 kep}^{-1}$
 \\
$t_{\rm ff, disc} =\sqrt{{1}/{G\rho_{\rm disc}}}$ & Local mean free-fall time of
the disc \\
$H_{\rm disc}$  & Disc scaleheight\\
$Q_{\rm disc}$  & Toomre $Q$ parameter of the disc\\
$Q_{\rm gas}$  & Toomre $Q$ parameter of the gas component of the disc\\

$f_{\rm H}$ & Disc scaleheight correction due to self-gravity and
gravitational force from the stars, $f_{\rm H}\approx 1$\\
 $D$ & Dissipation parameter (proposed in this paper), dissipative systems have
 $D > 1$, clump formation requires $D>1/3$\\
   \hline			

\end{tabular}
\caption{List of variable definitions. The microscopic
variables are the variables defined on the cloud scale,
and the macroscopic variables are the variables defined on
the disc scale. \label{tbl:variables}}
\end{table*}

\section{The model}
We describe the properties of a system consists of coagulating
clouds, and derive the time-scale for cloud-cloud collisions to 
 significantly remove kinetic energy from the system. We evaluate the
 stability of galactic discs that consist of coagulating clouds, and show that when the  mean surface
density of a disc is comparable or larger than the mean surface density of
molecular clouds in the disc, cloud collision cooling is effective. A list of variables can be found in table \ref{tbl:variables}.

\subsection{General picture}\label{sec:general}
In our colliding clouds picture, molecular clouds are
surrounded by a hot ambient medium. The clouds are mainly composed of H$_2$ gas,
and the ambient medium includes gas in both the warm neutral phase and the warm
ironized phase.

{ The structure formation in such a medium takes several steps. First, one
needs to create molecular gas, and second, one needs to assemble the molecular
gas into structures on different scales.  We will be focusing on the
structure formation, and therefore  we assume that there is a continuous of
supply of small clouds which eventually coagulate to form larger clouds.
 In reality, there are several ways to form these small clouds:
If the cold gas from out of the dynamically-triggered thermal instability, the
structure formation is likely to occur in a bottom-up way, with smaller
structures being created first and larger structures being built through the
coagulation of smaller structures. This is largely caused by
the shape of ISM cooling curve, and details concerning this can be found in
\citet{2016arXiv161001164M}.  Another mechanism to produce small clouds is
stellar feedback, which disrupts larger clouds into smaller ones. Given these
possibilities, we are able to assume the existence of a population of clouds,
and focus on the growth of structures in such a
clumpy medium.
}

\begin{figure*}
\includegraphics[width=0.75\textwidth]{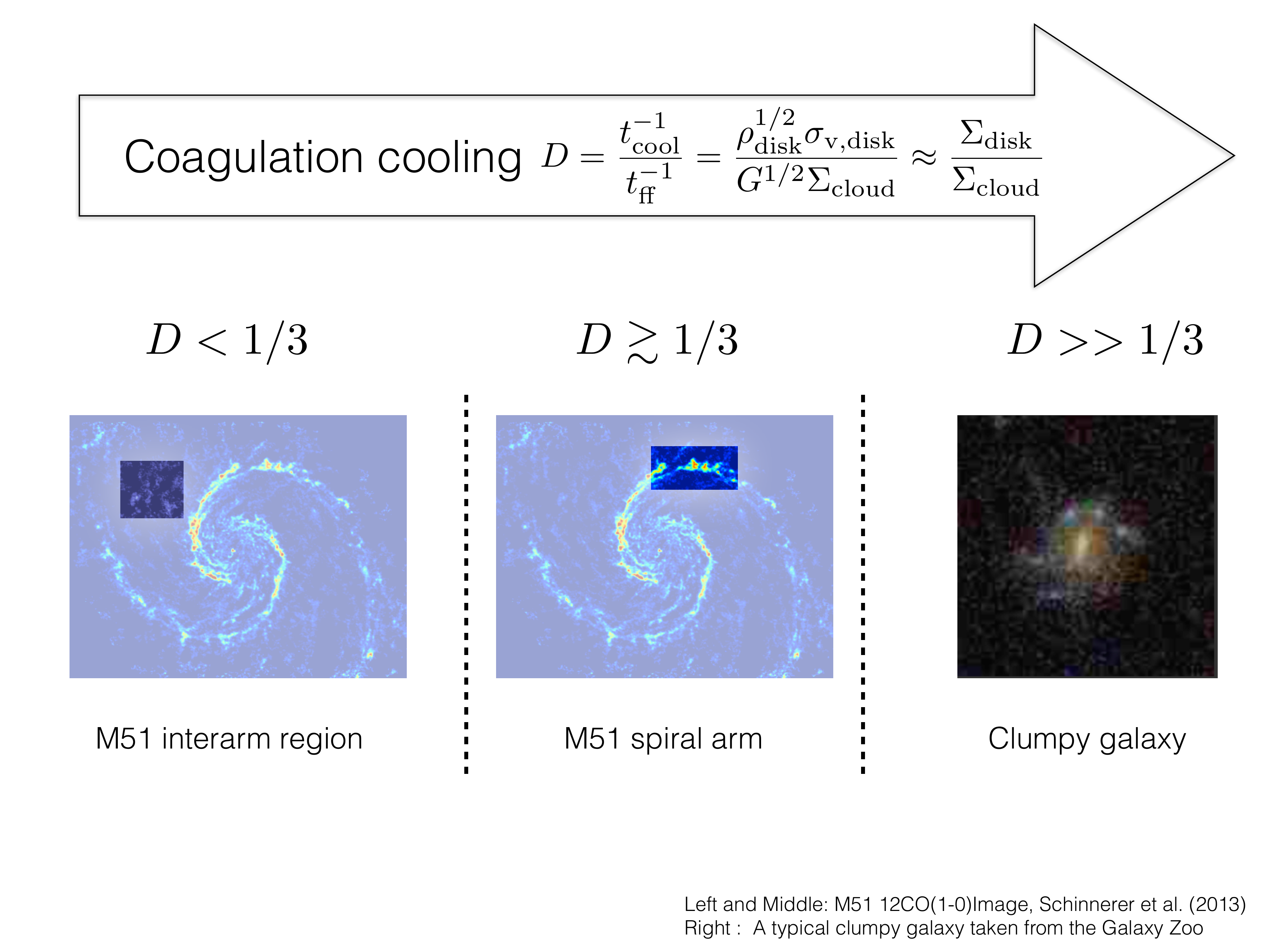}
\caption{\label{fig:3}  Different regimes of cloud collision cooling. In regions
like the M51 interarm region (see the highlighted box of the image on the left), 
cloud collision
cooling is relatively inefficient ($D < 1/3$, $t_{\rm cool} > 3 \Omega_{\rm
kep}^{-1}$, which correspond to $\Sigma_{\rm  disc} < 1/3 \Sigma_{\rm cloud}$),
and larger clouds are built by slowly accreting smaller clouds.
In the cooling-efficient regimes (middle and right panels), $D > 1/3$, $t_{\rm cool}
> 3 \Omega_{\rm kep}^{-1}$ (which also correspond to $\Sigma_{\rm  disc} <
1/3 \Sigma_{\rm cloud}$), cloud collision cooling can effectively remove
kinetic energy from the disc within a dynamical time, which enables the disc to
contract and form complexes seen on the spiral arms of the M51 galaxy (see the
highlighted box of the image in the middle) and kpc-size
giant clumps seen in gas-rich galaxies (e.g. the clumps seen in the galaxy on
the right).
Image credit:
The images of M51 galaxy are reproduced from the
velocity-integrated $^{12}$CO(1-0) data presented in
\citet{2013ApJ...779...42S}. The image of the clumpy galaxy is reproduced from
the Galaxy Zoo project \citep{2012amld.book..213F}, where the data is
produced by the Sloan Digital Sky Survey \url{http://www.sdss.org}. The reuse of
the figure is granted by the Creative Commons Attribution-Noncommercial-No
Derivative Works 2.0 UK. See Sec. \ref{sec:obs} for details.}
\end{figure*}

{ Our next step is to study the structure formation.
In this paper, the term cloud refers to  single, centrally condensed structures
whose sizes are smaller than the disc scaleheight, and more complex structures
whose sizes are comparable to the disc scaleheight are called complexes (in
Milky Way-like galaxies) and clumps (in gas-rich discs), respectively.
 Observationally, clouds appear as close-to-isolated objects, and complexes and
 clumps contain a significant amount of sub-structures 
\citep[e.g. see][for the case of the clumps]{2017MNRAS.464..491F}
and \citet[][for the case of a complex]{2011A&A...529A..41N}.\footnote{We must
acknowledge that it is still difficult to come us with more rigorous definitions to these objects. For example, many of the complexes in the Milky
Way are also called clouds in some papers. Here, we make distinction since
observations do indicate that clouds can organise themselves on larger scales.
An example of this can be found in the M51 galaxy
\citep{2008MNRAS.391..844D}, see also our Fig. \ref{fig:3}.} Since the
clumps and complexes are more structured, we propose that they are built by
quickly assembling the clouds on larger scales, and in this paper we study this assembly process.

To understand how structures form, one needs to describe the
multi-phased structure of the ISM. We assume that the ISM of gas-rich discs can
be described as an ensemble of molecular clouds, and the properties of the ISM
on the large scale can be derived by considering the interactions between the
 molecular clouds.  
%  To understand our arguments one
% should look at the multi-phase structure of the ISM.
% The ISM is mainly heated by photons at UV, X-ray and gamma-ray band, and it
% cools down due to various emission lines, and emitting bands. The cooling of the
% ISM is caused by different cooling agents that can be excited at different
% temperatures, and this produces wiggle-like features in the cooling curve.
% Roughly speaking, the different phases correspond to different wiggles on the
% cooling curve \citep{1969ApJ...155L.149F,1977ApJ...218..148M}, and this is
% independent on either a galaxy is clumpy or not.
This is justified because we are interested in gas-rich discs, where more than
half of the gas is molecular \citep{2010Natur.463..781T}. In these discs, the majority of
the mass and hence the kinetic energy is contained in molecular gas.
Because the molecular gas has densities that are much higher \footnote{E.g. the
molecular gas a density of $>10^3 \rm cm^{-3}$, which is already $\sim$ 100
times larger than the density of the surrounding warm neutral medium.  Between
these two phases, there exists the 
so-called cold neutral medium which contains mostly cold neutral hydrogen
of a temperature of $\sim 100$ K, but this phase contains very
little mass, and is usually not dynamically important. } than the density
of the ambient medium, we can neglect the drag force between the molecular
clouds and the ambient medium \citep{1983Ap&SS..89..177M}, and threat the
molecular ISM as an ensemble of clouds that interact through cloud-cloud
collisions.}
% This has two consequences: First, the clouds are so dense that their movements
% are barely affected by the ambient medium (as has been estimated by
% \citet{1983Ap&SS..89..177M}). Second, because of the density contrast, it
% would be difficult for turbulence in the ambient medium to cascade into the
% clouds. Recent studies suggest taht turbulence in the molecular clouds is
% mainly driven by cloud cloud-cloud collisions, whereas turbulence in the
% ambient medium is driven by a combination of shear and supernova feedback
% \citep{2011ApJ...738..101G,2011MNRAS.411...65B,2016MNRAS.458.1671K,2016ApJ...824...41I,2017MNRAS.465..667L}.

\begin{figure}
\includegraphics[width=0.45 \textwidth]{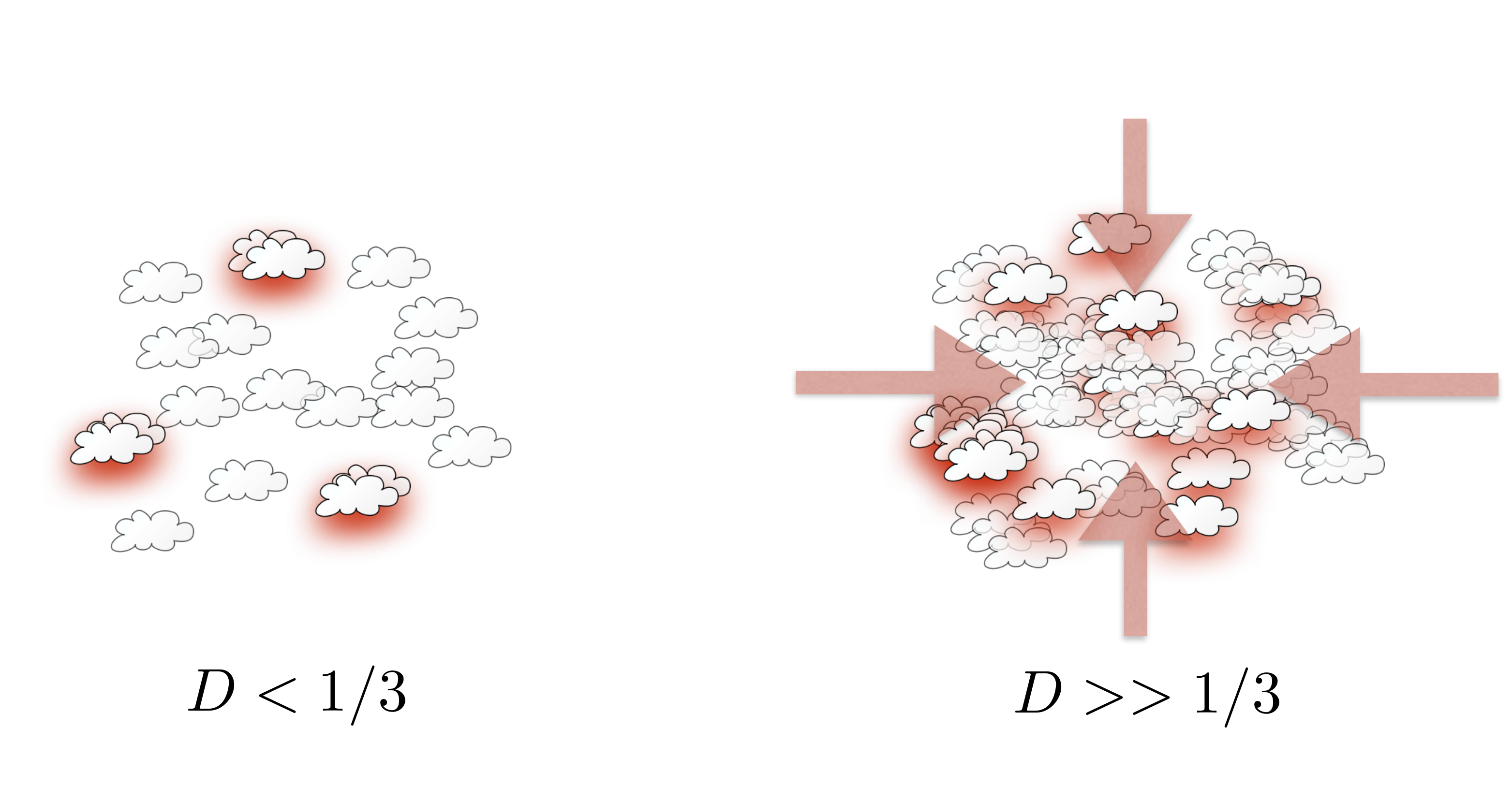}
\caption{ Illustration of different regimes of disc evolution. The left image
illustrate the almost-dissipationless case ($D<1/3$), where cloud-cloud collisions
occur at a relatively low frequency. The right image shows the dissipative case
($D>>1/3$), where cloud-cloud collisions occur at a high frequency. The
collisions lead to an effective energy dissipation. When the energy dissipation
time is shorter compared to the mean free-fall time of the system, cloud collision
cooling will cause
 the system to contract globally (as indicated by the
red arrows). This will drive the formation of clumps and complexes.
\label{fig:illus1} }
\end{figure}

{ 

 Based on this, we propose that to describe the dynamical properties of such
 a multi-phased medium, it is sufficient to construct a model that consists of
 an ensemble of molecular clouds, where the properties of the medium on the large
 scale can be determined by considering the collisions between the clouds. 
  In our view, the collisions of clouds have two major consequences on the disc evolution. The first and perhaps most obvious
 consequence of cloud-cloud collisions is that they lead to the formation of
 larger structures. This has been previously discussed
 \citep{1983ApJ...271..604K,1983Ap&SS..89..177M,1986PASJ...38...95T,2000ApJ...536..173T,2002A&A...382..872V,2008MNRAS.391..844D}. 
 Note that these collisions can produce clouds characterized by a power-law
 mass spectrum, similar to what is observed. But we argue that there is another
 consequence of cloud-cloud collisions,
 namely that these collisions should lead to a dissipation of kinetic energy in
 the disc, where the disc kinetic energy is converted into molecular cloud
 turbulence through these collisions.
 Following \citet{1989ApJ...344..306E}, we name this process cloud collision
 cooling.
 We propose that there are two regimes within which a galactic disc would evolve:
 there is almost-dissipationless regime, where larger clouds are built through the
gradual coagulation of smaller clouds. Since the clouds are well-separated, the
rate of cloud-cloud collisions is low. This regime is relevant for galaxies like
the Milky Way where the disc is less massive.
On top of this, we propose that there is a
new regime, where the rate of cloud-cloud
collisions is so high that the energy dissipation due to clouds-cloud collisions
has a non-negligible effect on the disc dynamics. In this dissipative regime,
cooling can effectively remove kinetic energy from the disc, and this loss of
kinetic energy will enable disc to contract form a larger scale where the clouds can coagulate into clumps.
As a result, each clump should contain a significant number of
overlapping structures reminiscent of the colliding clouds. The difference
between these two regimes is illustrated in Fig.
\ref{fig:illus1}. The paper is an investigation of the effect of cloud collision cooling on the evolution of galactic discs in different regimes.}

We do not explicitly consider the effect of stellar feedback. Together with
accretion, stellar feedback is a source of heating to the disc, which maintains
its velocity dispersion. Here, we simply assume that both
accretion and stellar feedback would drive the velocity dispersion of
the disc which ensures $Q_{\rm disc}\approx 1$, and we study the effect of cloud
collision cooling on the subsequent evolution of such a $Q_{\rm disc}\approx 1$
disc \footnote{For spiral galaxies, $Q_{\rm disc}$ should be computed by
combining the contributions from the gas and the stars
\citep{1992MNRAS.256..307R,1994ApJ...427..759W,2011ApJ...737...10E,2011MNRAS.416.1191R,2013MNRAS.433.1389R,2017arXiv170102138R}.
}.
 Since cloud collision cooling mechanism that we are discussing is
independent on how the clouds are created or disrupted, the link we
establish between cloud collision cooling and disc fragmentation is independent on the details of feedback and cloud destruction.
{ However, we should note that stellar
feedback is still important for our model as it disrupts bigger clouds
into smaller clouds, which helps to sustain a population of clouds of different
masses among which collisions should occur. }

\subsection{Disc cooling through cloud-cloud
collisions}\label{sec:coag:time}

To obtain a quantitative understanding of the underlying physics, we assume 
that a typical cloud has a mass $m_{\rm cloud}$ and a size $r_{\rm cloud}$.
{ In fact our results do not depend on the assumed the masses and sizes, but
only on the mass-weighted mean surface density of the cloud.} We further assume
that when averaged over a larger volume (whose size is much larger than the size of a molecular cloud, but is comparable to the disc scaleheight), the molecular galactic disc as a
whole has a local mean density of $\rho_{\rm disc}$ and velocity dispersion
$\sigma_{\rm v, disc}$. 

The number density of clouds is thus $n_{\rm cloud}
=\rho_{\rm disc}/ m_{\rm cloud}$. The mean free path of clouds in the disc is
\begin{equation}
\lambda_{\rm cloud} = \frac{n_{\rm cloud}^{-1}}{s_{\rm cloud}}= \rho_{\rm
disc}^{-1} m_{\rm cloud} r_{\rm cloud}^{-2}\;,
\end{equation}
where $s_{\rm cloud} \approx r_{\rm cloud}^2$ is the surface area of a
cloud .

Since cloud-cloud collisions are inelastic (such that each collision
removes a significant portion of the kinetic energy), the typical time for
the system to lose kinetic energy can be estimated from the typical time for a
cloud to collide with another cloud, which is
\begin{equation}\label{eq:tcoll}
t_{\rm cool}\approx t_{\rm collide} \approx  \lambda_{\rm cloud} / \sigma_{\rm
disc} = \rho_{\rm disc}^{-1} m_{\rm cloud} r_{\rm cloud}^{-2} \sigma_{\rm v,  disc}^{-1} =
\frac{\Sigma_{\rm cloud}}{\rho_{\rm disc}\sigma_{\rm v,  disc}}\;,
\end{equation}
where $\Sigma_{\rm cloud}$ is the mean surface density of the clouds \footnote{In the Milky Way, the velocity dispersion of the spiral arm at
$\sim 5\;\rm kpc$ is around 30 km/s, and mean density of the molecular disc is
$0.02\rm \; M_{\odot}\;\rm pc^{-3}$. Assuming that the clouds have 
$10 \; \rm M_{\rm \odot}\;\rm pc^{-2}$ (Sec. \ref{sec:sigma:cloud}), we derive a
cloud mean free path of $\sim 500\;\rm pc$, and a cloud collision time of $\sim
20 \;\rm Myr$. These values are taken from \citet{2006PASJ...58..847N}, and the
estimate cloud collision time agrees qualitatively with the value derived by
\citet{2015MNRAS.446.3608D}, which is  $\sim
10 \;\rm Myr$.}.
For comparison, the free-fall time of the gas is
\begin{equation}\label{eq:tff}
t_{\rm ff}\approx \sqrt{\frac{1}{G \rho_{\rm disc}}} \;.
\end{equation}
The collision time depends on both the bulk properties of the gas such as the
mean density $\rho_{\rm disc}$, the mean velocity dispersion $\sigma_{\rm
disc}$, as well as the properties of the individual clouds $m_{\rm cloud}$ , $r_{\rm
cloud}$. When the mean surface densities of the molecular clouds $\Sigma_{\rm
cloud}= m_{\rm cloud}/ r_{\rm cloud}^2$ and the velocity dispersion of the systems $\sigma_{\rm v, disc}$ are fixed,
the cloud collision cooling time
is proportional to $\rho_{\rm disc}^{-1}$, and the free-fall time is proportional to $\rho_{\rm
disc}^{-1/2}$, it is clear that cloud collision cooling will be important at
regions where the mean density of the disc $\rho_{\rm disc}$ is sufficiently
high.
Whether the disc is dissipative or not is quantified by the dissipation parameter
\begin{equation}\label{eq:d}
D = \frac{t_{\rm cool}^{-1}}{t_{\rm ff}^{-1}}=\frac{\rho_{\rm
disc}^{1/2}\sigma_{\rm v, disc}}{G^{1/2} \Sigma_{\rm cloud}}\;,
\end{equation}
 which is the ratio between the rate of
cloud collision cooling $t_{\rm cool}^{-1}$ and the rate of gravitational
collapse $t_{\rm ff}^{-1}$.
Systems with large $D$ are dissipative. Note that $D$ is only dependent on the
mean surface density of the cloud. Therefore, our equations are accurate when
the clouds in a galaxies share a common surface density. { When the clouds
have different surface densities,  since we are interested in the
dynamical properties of molecular gas on the large scale, $\Sigma_{\rm cloud}$
should be mass-weighted mean surface density the molecular clouds. We explain in
Sec. \ref{sec:sigma:cloud} that a constant $\Sigma_{\rm cloud}$ for  a galaxy is
likely a good assumption.}

\begin{figure}
\includegraphics[width=0.5\textwidth]{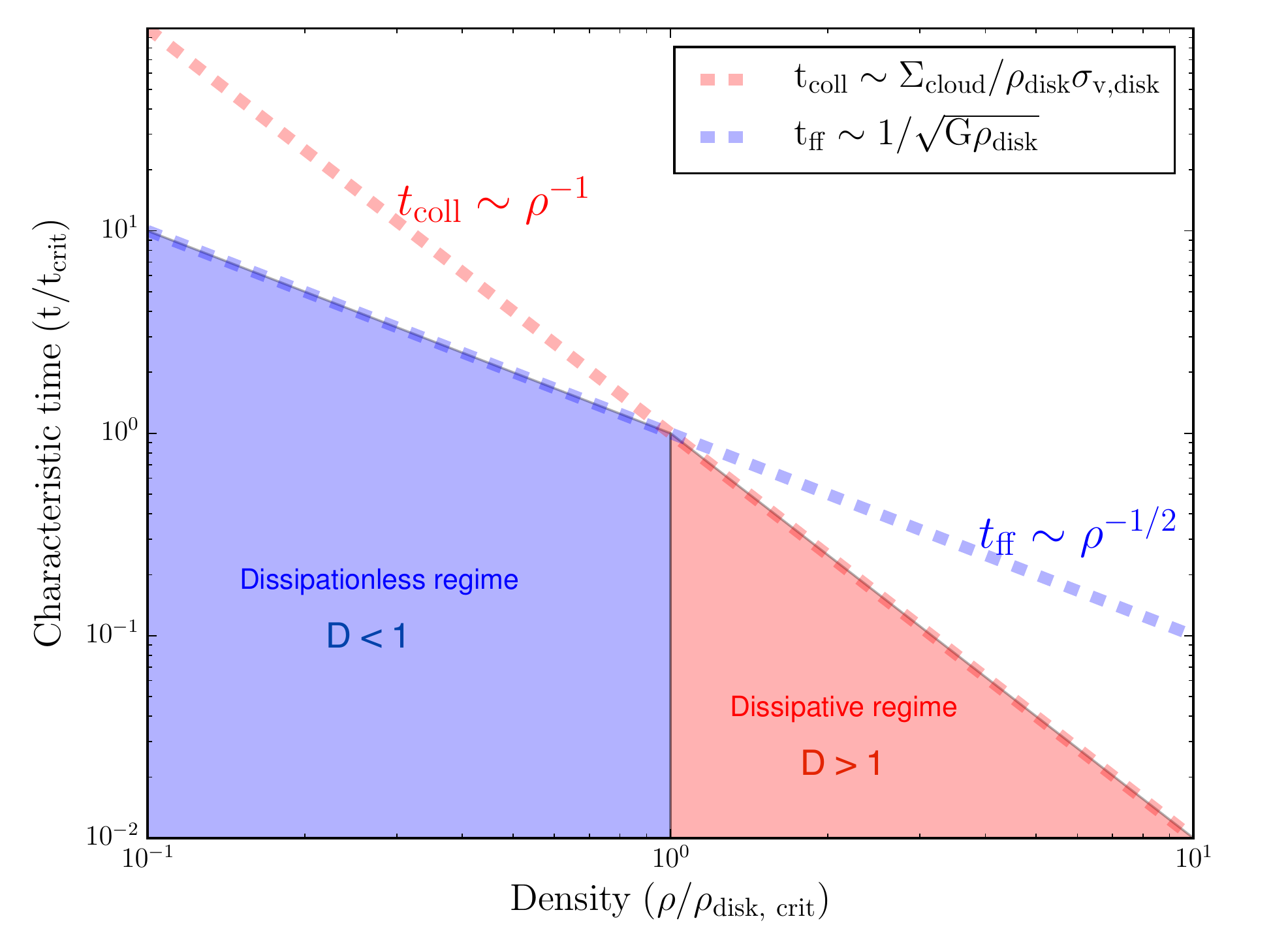}
\caption{Dynamical effect of cloud-cloud collisions on the evolution of the
galactic ISM.
The medium has a mean density of $\rho_{\rm disc}$, and a velocity
dispersion $\sigma_{\rm v, disc}$, and typical molecular clouds have a surface
density of $\Sigma_{\rm cloud}$. The mean free-fall time of the gas in the disc
is  $t_{\rm ff}\sim \rho_{\rm disc}^{-1/2}$, and the typical
time for cloud-cloud collisions to occur is $t_{\rm cool} \sim \rho_{\rm
disc}^{-1}$.
We quantify the system with the $D$ parameter, which is the ratio between the
rate of cloud collision cooling $t_{\rm cool}^{-1}$ and the rate of gravitational
collapse $t_{\rm ff}^{-1}$. When $D < 1$ (corresponding to $t_{\rm ff} < t_{\rm
cool}$, $\rho_{\rm disc} < \rho_{\rm
disc, crit}$), the system of gas clouds behaves as an almost-dissipationless gas, as
indicated by the blue shaded region, and when $D > 1$ (corresponding to $t_{\rm
ff} > t_{\rm cool}$, $\rho_{\rm disc} > \rho_{\rm disc, crit}$), the clouds
form a dissipative system and would lose a significant faction of
their kinetic energy within a free-fall time due to cloud-cloud collisions
(as indicated by the red shaded region). The density is normalised with respect
to $\rho_{\rm disc,\; crit}$ given by Eq.
\ref{eq:rhocrit}, and the time is normalised with respect to $t_{\rm crit} = 1/
\sqrt{G \rho_{\rm disc,\; crit}}$. 
\label{fig:gas} }
\end{figure}
Fig. \ref{fig:gas} plots the cloud collision time and the free-fall time as a
function of mean density $\rho_{\rm disc}$. One can identify these two
regimes:
when the density is low, the free-fall time is shorter than the characteristic
time for two clouds to collide, $D <1$, and the kinetic energy contained in the
system is roughly conserved in a free-fall time.
When the density is large enough e.g.
\begin{equation}\label{eq:rhocrit}
\rho_{\rm disc} > \rho_{\rm disc, crit} = \frac{G \Sigma_{\rm
cloud}^2}{\sigma_{\rm v,  disc}^2}\;,
\end{equation}
$D> 1$, 
 collisions between the clouds can effectively remove the kinetic energy from
 the system within a dynamical time, and the system is dissipative. If such a system
has a kinetic energy density of $e_{\rm kin}$, the cooling rate can be
estimated via 
\begin{equation}\label{eq:diss:1}
\dot e_{\rm kin}\approx e_{\rm kin}/t_{\rm cool}\;.
\end{equation}
Therefore, collisions between molecular clouds
that occur on the microscopic scale  can reduce the velocity dispersion of the system, and
constitute as a macroscopic, dynamical cooling channel. This cooling
channel is different from radiative cooling, which merely removes the
thermal energy but not the kinetic energy. Following
\citet{1989ApJ...344..306E}\footnote{Where he also derived the dispersion
relation of such a clumpy medium.}, we name this dynamical cooling process
``cloud collision cooling''.

  One might wonder the effect of cloud destruction
due to stellar feedback:
 what would happen if the cloud is destroyed by stellar feedback
 before it can collide with another cloud? Although the finite lifetime of molecular
 clouds is an importance factor to consider when one wishes to build an
 intuitive picture of disc evolution, whether the cloud lifetime is longer or
 shorter than the cloud collision time does not affect our cooling rate
 estimate.
 This is because our cloud collision cooling mechanism works as long as the
 collisions lead to energy dissipation, and this is independent on if the clouds
 can survive between two successive encounters or not.

\begin{table*}
\begin{center}
\begin{minipage}{\linewidth}
\small
\centering
\begin{tabular}{l | r | r | r |}
  \hline

Region  &$\Sigma_{\rm  disc}$ 
&$\Sigma_{\rm cloud}$ &  $ D =  \Sigma_{\rm  disc}/ \Sigma_{\rm cloud}$ 
\\
  \hline	
Milky Way interarm region &  $1  M_\odot\;\rm pc^{-2}$ &  $ 12 M_{\odot}\;\rm
pc^{-2}$  & 0.1 \\
Milky Way molecular ring  & $ 10 M_\odot\;\rm pc^{-2}$&   $ 12 M_{\odot}\;\rm
pc^{-2}$ & 1  	\\
 M51 Interarm region  &  $ 4 M_\odot\;\rm pc^{-2}$&   $ 12 M_{\odot}\;\rm
 pc^{-2}$ & 0.3
 \\
 M51 outer spiral arm  &  $ 15 M_\odot\;\rm pc^{-2}$ &  $  12M_{\odot}\;\rm
 pc^{-2}$ & 1 \\
Giant clumps in gas-rich galaxies  & $ 100 M_\odot\;\rm pc^{-2}$& 
$ 30 M_{\odot}\;\rm pc^{-2}$ & 3	\\
\end{tabular}
\caption{A list of the estimated parameters for different galactic environments.
\label{tbl:regions}}
\vspace{0.1 cm}
\end{minipage}\\
%\end{center}
Notes: The Milky Way values are measured from data presented in
\citet{2006PASJ...58..847N} where we adopt a galactocentric distance of 5
kpc. The interarm region and spiral arm region of the M51 galaxy are
measured at a galactocentric distance of 5 kpc, from data
presented in \citet{2007A&A...461..143S,2009A&A...495..795H}. 
To estimate the properties of discs hosting
massive clumps, we adopt typical values from the literature
\citep{2005ApJ...627..632E,2011ApJ...733..101G,2012MNRAS.422.3339W,2015ApJ...806L..17S}.
We adopt a typical size of $\sim 1 \;\rm kpc$ and a typical mass of $10^9\;
M_{\odot}$, from which we derive a surface density of $10^3 \;M_{\odot}\;\rm pc^{-2}$. 
We estimate that giant clumps occupy $\sim 10\%$ of the
total disc surface, and the mean surface density of the disc is  estimated to be
$10^3 \;M_{\odot}\;\rm pc^{-2}\times 10\% = 10^2 M_{\odot}\;\rm pc^{-2}$.
The values of $\Sigma_{\rm cloud}$ are computed using Eq. \ref{eq:sigma:cloud},
where we have assumed solar metallicity for the Milky Way and M51, and assumed
0.4 solar metallicity for clumpy galaxies. Forming clumps requires $D>1/3$,
which corresponds to  $ \Sigma_{\rm  disc}/ \Sigma_{\rm cloud}>1/3$. 
\end{center}
\end{table*}

\subsection{Impact of cooling on disc fragmentation}\label{sec:cooling}
The ratio between cooling time and dynamical time determines disc evolution:
\citet{2001ApJ...553..174G,2003ApJ...597..131J} simulated the fragmentation of self-gravitating gas discs, and demonstrated that the
evolution of the disc in the nonlinear regime is determined by the ratio between
the cooling time $t_{\rm cool}$ and $3\Omega_{\rm kep}^{-1}$
(which is roughly the disc dynamical time $t_{\rm kep}$).
  The cooling time plays a determining role in
 the dynamics of the disc: when
 $t_{\rm cool} < 3\Omega_{\rm kep}^{-1}\approx t_{\rm kep} $, cooling is not
 significant,  the disc evolves into a gravo-turbulent state where gas forms long, filamentary
 structures that are constantly sheared apart by disc rotation; when $t_{\rm
 cool} < 3\Omega_{\rm kep}^{-1}\approx t_{\rm kep}$, cooling is important, and the disc quickly fragments into clumpy gas
condensations.

% 
% We consider the effect of cloud coagulation on disc dynamics.
% In our model, the energy dissipation takes two steps: in the first step,
% kinetic energy of the disc is converted into the turbulent motions in the
% individual molecular clouds due to cloud coagulation. This process effectively
% removes kinetic energy from the disc, and it leads to cooling and enables disc
% fragmentation.
% In the second step, turbulent energy in the individual clouds is dissipated
% into heat.
% The energy conversion is further discussed in Sec. \ref{sec:balance}.

The cooling time of the disc on the macroscopic scale is comparable to the cloud
collision time on the microscopic scale, which is (see Eq. \ref{eq:tcoll})
\begin{equation}\label{eq:t:coal}
t_{\rm cool} \approx \frac{\Sigma_{\rm cloud}}{\rho_{\rm disc}
\sigma_{\rm v,  disc}}\;,
\end{equation}

We can further simply the expression of $t_{\rm cool}$ using some additional
ansatz: When the disc reaches hydrostatic equilibrium along the vertical
direction, we can estimate its scaleheight (see Appendix. \ref{sec:h})
\begin{equation}\label{eq:H}
H_{\rm disc} = \frac{f_{\rm H} \sigma_{\rm v,  disc}}{\Omega_{\rm kep}}\;,
\end{equation}
 where $f_{\rm H} \approx 1$ is a
 numerical factor taking into account the additional compression contributed
 from self-gravity and the gravitational force from the stars. For order-of-magnitude analysis, we assume $f_{\rm H}\approx 1$. 
 Combining Eq.
\ref{eq:t:coal} and Eq. \ref{eq:H}:
\begin{equation}\label{eq:t:cool:new}
t_{\rm cool} = f_{\rm H} \frac{\Sigma_{\rm cloud}}{\Sigma_{\rm disc}} \times
\frac{1}{\Omega_{\rm kep}} \approx \frac{\Sigma_{\rm cloud}}{\Sigma_{\rm disc}}
\times \frac{1}{\Omega_{\rm kep}}\;.
\end{equation}

According to \citet{2001ApJ...553..174G}, to decide if a disc would fragment or
not, one needs to compare the cooling time $t_{\rm cool}$ with the disc
dynamical time $t_{\rm kep} \approx 3 \Omega_{\rm kep}^{-1}$. However, 
for a disc that is self-regulating, $Q_{\rm disc}\approx 1$, the local free-fall
time of the disc gas is always comparable to the disc dynamical time ($t_{\rm
ff}\approx 1/3 t_{\rm kep}\approx  \Omega_{\rm kep}^{-1}$) \footnote{The
free-fall time is $t_{\rm ff}\approx 1/\sqrt{G\rho_{\rm disc}}$, and the disc
dynamical time is $t_{\rm kep} \approx 1/\Omega_{\rm kep}$. Using Eq.
\ref{eq:H}, $\rho_{\rm disc}\approx \Sigma_{\rm disc}/
H_{\rm disc} \approx  \Omega_{\rm kep} \Sigma_{\rm
disc} /\sigma_{\rm v, disc} \approx G^{-1} \Omega_{\rm kep}^2 /Q_{\rm disc }$.
Since $Q_{\rm disc}\approx 1$,
$t_{\rm ff}\approx 1/\sqrt{G\rho_{\rm disc}} \approx Q_{\rm disc}^{1/3}
\Omega_{\rm kep}^{-1}\approx 1/3 t_{\rm kep}$ .}.
Thus, instead of evaluating the ratio between $t_{\rm cool}$ and $t_{\rm kep}
\approx 3 \Omega_{\rm kep}^{-1}$, one can also compare the $t_{\rm cool}$ with
the local free-fall time $t_{\rm ff}$ to decide the disc fragmentation.
In fact, the ratio between cooling time $t_{\rm cool}$ and the local free-fall
time $t_{\rm ff}$ is related to the $D$ parameter that we have derived before
($t_{\rm cool}/3 \Omega_{\rm kep} \approx t_{\rm cool}/3 t_{\rm ff} = 1/3
D$)! In other words, $D$ can also be used to decide if a disc will fragment into
clumps, and clump-forming discs have $D > 1/3$.
Using Eq. \ref{eq:t:cool:new},
\begin{equation}
D = \frac{t_{\rm cool}^{-1}}{t_{\rm ff}^{-1}} = \frac{\Sigma_{\rm
disc}}{f_{\rm H} \Sigma_{\rm cloud}}\;,
\end{equation}
where it is simply determined by the ratio between  the mean surface density of
the galactic disc $\Sigma_{\rm  disc}$ and the mean surface density of molecular
clouds in the disc $\Sigma_{\rm cloud}$. Fragmenting discs should have $D > 1/3$
(therefore $\Sigma_{\rm disc}> 1/3 \Sigma_{\rm cloud}$).

\subsection{Surface densities of clouds}\label{sec:sigma:cloud}

Our next goal is to determine the ``typical'' surface density of
the clouds. In a galaxy, the clouds can have different masses and sizes. The
typical surface density is determined from the mass-weighted mean of the
surface densities of the individual clouds. We argue that this mean value is
mainly determined by the metallicity of a galaxy.

Our argument is based on the analysis by 
\citet{2008ApJ...689..865K,2009ApJ...693..216K} where they considered the
atomic-to-molecular transition and they self-consistently modelled H$_2$
formation, star formation, photodissociation and shielding. They found that the
limiting surface density of atomic-to-molecular transition is determined mainly by the metallicity, and is only weakly dependent on the interstellar
radiation field. Besides, the fact that a cold phase must exist also provides
addition constraints to the strength of the interstellar radiation field. As a
result, the critical surface density for H\,{\sc i}/H$_2$ transition is merely a
function of  the metallicity.
Following them, \citet{2014ApJ...790...10S} found that the
following formula can provide a reasonably good description to the observed
H\,{\sc i}/H$_2$ transition at Galactic and extragalactic observations
\citep[e.g.][]{2012ApJ...748...75L,2017ApJ...835..126B}
 \footnote{Some
observations indicate higher values of the typical surface density of the
clouds.
For the Milky Way, the reported surface density of the
clouds ranges from $40 M_{\odot}\;\rm pc^{-2}$
\citep{2009ApJ...699.1092H} to 
$200 M_{\odot}\;\rm pc^{-2}$ \citep{2010ApJ...723..492R}. As
\citet{2010ApJ...723..492R,2015ARA&A..53..583H} noted, the major uncertainty
comes from the different ways to define the cloud boundary, and these values should be
considered as upper limits.
These surveys also miss a significant fraction of the total mass due to source extraction
\citep{1989ApJ...339..919S}. Because of these, the values derived from
the dedicated studies such as \citet{2012ApJ...748...75L,2017ApJ...835..126B}
are more reliable.}:
\begin{equation}\label{eq:sigma:cloud}
\Sigma_{\rm cloud, crit} \approx  \frac{12}{Z}\; M_{\odot}\;{\rm pc}^{-2}\;,
\end{equation}
where $Z$ is measured in terms of solar abundance.

Eq. \ref{eq:sigma:cloud} predicts the typical surface density a cloud should
have in a galaxy. If a cloud has a surface density that is lower than the value
predicted by Eq. \ref{eq:sigma:cloud}, it is vulnerable to photodissociation.
On the other hand, if a cloud has a much higher surface density, star formation in such
a cloud would destroy the cloud within a relatively short time, making these
high surface density clouds rare. In fact, as \citet{2010ApJ...710L.142F} pointed
out, most of the clouds in a galaxy should have surface densities that are 
not much higher than the value predicted by Eq.
\ref{eq:sigma:cloud}.

One can look for evidence of this almost-constant surface density from
observations: in the Milky Way, it is relatively easy to measure the surface
density of the well-resolved clouds, and these observations do indicate that
the HI/H$_2$ occurs at the typical surface density predicted by Eq.
\ref{eq:sigma:cloud}. Recently,  \citet{2017ApJ...835..126B}  carried out a
similar study on the W43 molecular complex. This is an extremely complicated
region, where the gas is already evolved. They found that the total measured
surface density of can be separated into contributions from a few velocity components, and on average, each component has a surface density that is close to
the $12 M_{\odot}\;\rm pc^{-2}$, agreeing well with Eq. \ref{eq:sigma:cloud}. The still limited
observational evidences seem to support the idea that the clouds do stay close
to the threshold predicted by Eq. \ref{eq:sigma:cloud}.

For gas-rich discs, since the individual clouds are not resolved, it
remains an assumption that Eq. \ref{eq:sigma:cloud} predicts the characteristic
cloud surface density. We believe in this assumption since the physical
arguments are still valid in these environment. In practise, even if the cloud surface
density deviate from the  prediction of Eq. \ref{eq:sigma:cloud}, it is still
acceptable since we  are
mainly interested in comparing Milky-Way-like galaxies with gas-rich clumpy
galaxies whose surface densities are two orders of magnitudes higher. Compared to this variation, the deviation of the surface density of real clouds from the
one predicted by Eq. \ref{eq:sigma:cloud} is probably negligible.

 We should also emphasize that since clumps
 are objects that are formed from the disc instability enabled by cloud
 collision cooling where the clouds are allowed to organized on a larger scales,
 we do not expect Eq. \ref{eq:sigma:cloud} to hold for the clumps. Instead,
 since each clump contains a significant number of overlapping clouds,
 we expect $\Sigma_{\rm cloud}>> \Sigma_{\rm cloud}$.

\subsection{Condition of clump formation}
Based on the previous discussions, we  rewrite $D$ parameter as
\begin{equation}\label{eq:d:z}
D = \frac{t_{\rm cool}^{-1}}{t_{\rm ff}^{-1}}  =\frac{\rho_{\rm
disc}^{1/2}\sigma_{\rm v, disc}}{G^{1/2} \Sigma_{\rm cloud}}\approx \frac{\Sigma_{\rm
disc}}{f_{\rm H} \Sigma_{\rm cloud}} \approx \frac{Z}{12  f_{\rm H}}
\frac{\Sigma_{\rm disc}}{M_{\odot}\;\rm pc^{-2}}\;,
\end{equation}
where in the last step we have used Eq. \ref{eq:sigma:cloud}. { In practise,
one can express the $D$ parameter with a few different expressions: $D=\rho_{\rm
disc}^{1/2}\sigma_{\rm v, disc}/G^{1/2} \Sigma_{\rm cloud} $ is a rigours
definition, and $D \approx \Sigma_{\rm disc } / \Sigma_{\rm cloud}$ is
an approximate expression assuming that the disc reaches hydrostatic balance
along the vertical direction. In converting $D$ into the approximate
expression, we have assumed that the disc reaches hydrostatic equilibrium in the
vertical dimension and assumed that the gravitational force from the stars are
not significant. In this paper, we will be mainly using the approximate
expression more frequently as it is more convent, and a comment on the errors can be found in Sec.
\ref{sec:cooling} and Appendix \ref{sec:h}.

}
% 
% Practically speaking, $\Sigma_{\rm cloud}$ is
% determined by the metallicity of galaxies, which do not evolve significantly
% over the cosmic time. In contrast to this, $\Sigma_{\rm disc}$ can
% evolve by around two orders of magnitudes (see Table \ref{tbl:regions}).
% Therefore, the evolution of the $D$ parameter is largely determined by
% variations in $\Sigma_{\rm disc}$.

The evolution of a disc is therefore determined by the $D$ parameter, which is
related to the
ratio between the surface density of a galactic disc $\Sigma_{\rm disc}$ and
the typical surface density of molecular clouds in the disc $\Sigma_{\rm
cloud}$. There are these distinct regimes of disc evolution:
 When
$D < 1/3$, cloud collision cooling is inefficient, and gas is stretched into
filamentary structures due to shear. When $D>> 1/3$, cloud collision
cooling is efficient, and the disc fragments into clumps. 

What occurs in these different regimes?
When $D < 1/3$, the clouds do collide, but the rate of the collision is relatively
low, and energy dissipation from the cloud-cloud collisions is negligible. Thus
the impact of cloud collision cooling on the disc dynamics is relatively
insignificant. 
When $D>> 1/3$, $\Sigma_{\rm disc} >> \Sigma_{\rm cloud}$, { one should see
clouds overlap along the individual line of sights \footnote{To help with the
imagination, one is invited to consider a picture where $\rho_{\rm cloud} >
\rho_{\rm disc}$. However, since disc is relatively thick, the clouds are
overlapping along the individual line of sight, such that $\Sigma_{\rm disc} >
\Sigma_{\rm cloud}$. }.
} In this regime, clouds should collide with each other at a high frequency, and  cloud-cloud collisions can
cool down the disc within a dynamical time. The dissipation of kinetic energy
leads the disc to contract, and this drives the collective motion
of the clouds that leads to the formation of the giant clumps.
Therefore, what distinguishes a clumpy galaxies with normal spiral galaxies is
the value of the $D$ parameter.

% A normal, spiral galaxy does not fragment into
% clumps since the cloud-cloud collisions in these galaxies are not yet able to
% cool down the disc within a dynamical time, and in clumpy galaxies, the
% cooling due to cloud-cloud collisions is efficient which enables the disc to
% fragment.s

{In this paper, we are
interested in the effect of cloud collision cooling on the formation of
structure, and to fulfil our purpose $\Sigma_{\rm disc}$ should be the mean
surface density of the disc before fragmentation occurs.
For gas-rich discs, to evaluate the $D$ parameter, one needs to make an average
across the whole disc or over different radii. We are also
interested in the formation of regular-spaced molecular complexes in spiral galaxies. In this case, one should
 evaluate the $D$ parameter using the mean surface density of gas measured along
 the the spiral arms.}

Note that the dissipation parameter $D$ is proportional to the
metallicity (Eq. \ref{eq:d:z}). 
In other words, galaxies
with higher metallicities will dissipate their kinetic energy faster
compared to galaxies with lower metallicities. This can be understood as
follows:
when one increases the metallicity, the molecular
clouds require a lower surface density to shield
against the background radiation field, the clouds are thus less condensed. They
have lower surface densities, and are more like to collide with each other. This
leads to a higher rate of cloud-cloud collisions and a 
more efficient disc dynamical cooling.

\subsection{Kinetic energy budged and driving of molecular cloud
turbulence}\label{sec:balance} 
\begin{figure}
\includegraphics[width=0.45 \textwidth]{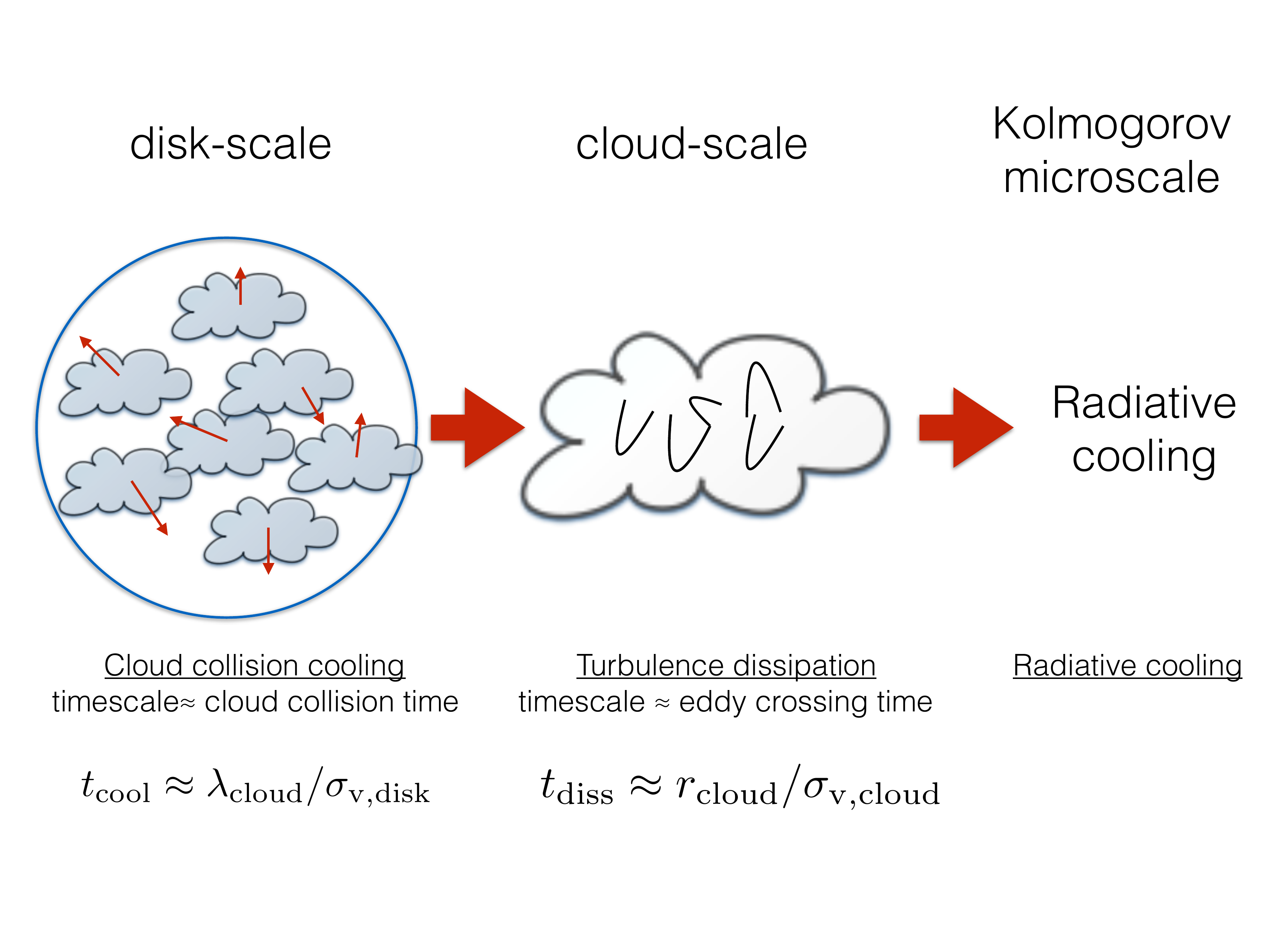}
\caption{\label{fig:cooling} A illustration of the proposed energy flow. On the
disc scale, the energy dissipation is dominated by cloud collision cooling. The
cooling time is  $t_{\rm cool} \approx \lambda_{\rm
cloud}/\sigma_{\rm v, disc}$ where $\lambda_{\rm cloud}$ is the mean free path of a molecular cloud in
the disc, and $\sigma_{\rm v,
disc}$ is the velocity dispersion of the disc. Cloud collision cooling converts
the kinetic energy of the disc into turbulence in molecular clouds. On the cloud
scale, the major cooling mechanism is the turbulence energy dissipation, which
converts the turbulence energy into heat. The characteristic time-scale of
turbulence dissipation is $t_{\rm diss}\approx r_{\rm
cloud}/\sigma_{\rm v, cloud}$ where $r_{\rm cloud}$ is the size of the cloud
and $\sigma_{\rm v, cloud}$ is the velocity
dispersion of a cloud.
Turbulence finally converts kinetic energy of the clouds into heat. On the
Kolmogorov microscale where turbulence cascades stops, the kinetic energy is
dumped into heat that is dissipated due to radiative cooling.
}
\end{figure}

In previous models of galactic discs
\citep[e.g.][]{2010MNRAS.407.1223R,2010ApJ...712..294E,2010A&A...520A..17K,2011ApJ...737...10E,2014MNRAS.438.1552F},
it has been assumed that the kinetic
energy in the disc can be efficiently dissipated into heat due to turbulence
dissipation. In our model, the
kinetic energy of the disc is contained in the random motions of the clouds, and
it is converted into molecular cloud turbulence
due to cloud-cloud collisions. In the final step, molecular cloud
turbulence is dissipated into heat.
The cloud-cloud collisions become a first step in the whole energy
conversion process, followed by turbulence dissipation. This is illustrated in
Fig. \ref{fig:cooling}. { Can we explain the observed level of molecular
cloud turbulence with cloud-cloud collisions?} As an estimate, the
energy dissipation rate  (of unit $\rm erg\; g^{-1}\; s^{-1}$) of a galactic
disc due to cloud-cloud collisions is (Eq. \ref{eq:diss:1})
\begin{equation}
\dot e_{\rm disc} = \frac{\sigma_{\rm v,  disc}^2}{t_{\rm cool}} \approx
\frac{\rho_{\rm disc} \sigma_{\rm v,  disc}^3}{\Sigma_{\rm cloud}}\;.
\end{equation}

Molecular
clouds in the Milky Way and presumably other galaxies are characterized by the
Larson relation \citep{1981MNRAS.194..809L,2004RvMP...76..125M} where, roughly
speaking $\sigma_{\rm v, cloud}\sim l^{1/3}$ where $\sigma_{\rm v}$ is the
velocity dispersion of a molecular cloud and $l$ is the size.
Assuming the kinetic energy of the clouds are dissipated due to turbulence, one
can determine the normalisation of the Larson relation:
$\sigma_{\rm v}\sim \dot e_{\rm cloud}^{1/3} l^{1/3}$ where
$\dot e_{\rm cloud}$ is the mean energy dissipation rate of molecular cloud turbulence
\footnote{Here we are interested in the normalisation of the Larson relation,
and do not consider the effect of fractal-like underlying density structures on
the slope of the Larson relation. Details concerning this can be found in 
 \citet{2007ApJ...665..416K}.}.
Assuming that the amount of energy a molecular cloud receives due to cloud-cloud
collisions can be effectively converted to turbulence, $\dot e_{\rm cloud}
\approx \dot e_{\rm disc}$, thus
\begin{equation}\label{eq:energy:convert}
\sigma_{\rm v, cloud}  \approx \sigma_{\rm v,  disc} \big{(}\frac{\rho_{\rm
disc}}{\Sigma_{\rm cloud}}\big{)}^{1/3}  r_{\rm cloud}^{1/3} \approx
\sigma_{\rm v, disc} \big{(}\frac{\Sigma_{\rm disc}}{\Sigma_{\rm
cloud}}\big{)}^{1/3} \big{(}\frac{r_{\rm cloud}}{H_{\rm disc}}\big{)}^{1/3}\;,
\end{equation}
where $H_{\rm disc}$ is the disc scaleheight. If cloud-cloud collisions
constitute as the major driver of molecular cloud turbulence, the level of
turbulence seen in molecular clouds is linked to the properties of the galactic discs on the
large scale! For discs of gas-rich galaxies where the surface densities are
high, we expect higher normalisations of the Larson's relation due to the
increased cloud-cloud collision rate.  Indeed, we do observe such an
increase of turbulence in gas-rich disc \citep[e.g.][]{2011ApJ...742...11S}. Although the
authors have attributed the cause of this increased level of turbulence to the
high ambient pressure, according to our model, it can also be explained by an
increase rate of cloud-cloud collisions.  We note that a similar concept has
been previously proposed by \citet{2013IAUS..292...19T} (followed by
\citet{2017arXiv170603764L}) where they emphasized that collisions inject
momentum into the clouds, and here we have made a different assumption where
collisions are mainly injecting the kinetic energy which then dissipates due to turbulence. We plan
to further investigate this issue in a forthcoming paper.

\section{Connection to observations}\label{sec:obs}

According to our analysis, whether a disc will fragment into clumps or not is
determined by the $D$ parameter.
  To estimate $D$,
we collect data from the literature, and the collected values are listed in
Table \ref{tbl:regions}. In our analysis, the values of $\Sigma_{\rm cloud}$ are
estimated using Eq. \ref{eq:sigma:cloud}, and arguments for this can be found in
Sec.
\ref{sec:sigma:cloud}. How $\Sigma_{\rm disc}$ is estimated is dependent on
the question we are interested in:
for clumpy galaxies, we are interested in the fragmentation of the disc, and the
value of $\Sigma_{\rm disc}$ is estimated by averaging over the whole disc. For spiral galaxies, since we are
interested in the fragmentation of different parts, the values  of $\Sigma_{\rm
disc}$ are evaluated by averaging over the spiral arms and the interarm
regions, respectively. 
% Through cosmic times, the metallicity of galaxies are determined by the
% interplay between cosmological gas inflow and metal production due to star
% formation, and it does not evolve by much. On the other
% hand, the surface density of the discs $\Sigma_{\rm  disc}$ evolve by a factor
% of $\sim 100$. Practically speaking, it is the variation of $\Sigma_{\rm disc}$
% that drives the galaxies to evolve differently, where clumps appears
% preferentially in gas-rich discs.

% 
% To look at interplay between cloud collision cooling and disc dynamics in
% different galaxies. We adopt the following terms: 
% \begin{itemize}
%   \item Cloud: the term ``cloud'' refers to typical Milky-Way
% molecular clouds where for each object there is a dominant emitting structure. The sizes of the clouds should
% also be smaller than a fraction of the disc scaleheight.
% \item Complex: the term ``complex'' refers to larger, centrally condensed
% structures where they are allowed to have multiple emitting peaks. However, different from the clumps,
% they reside in the close vicinity of the spiral arms. 
% \item Clump: the term ``clump'' is
% reserved for clumps in gas-rich discs where the sizes of the structure
% are comparable or larger than disc scaleheight.
% \end{itemize}
% In our picture,  clouds are objects that are ubiquitous in different
% galaxies.
% When $D$ is larger ($D\gtrsim 1$), they
% organise into complexes that stay along the spiral arms, and when $D>> 1/3$, they
% form clumps.
% These different regimes are summarized in Fig.
% \ref{fig:3}.
% 

For gas in interarm
region of the Milky Way and M51,  $D < 1/3$ (which roughly
corresponds to $\Sigma_{\rm disc} < 1/3\; \Sigma_{\rm cloud}$), cloud collision
cooling is inefficient (where, by definition, the time for a cloud to collide with another cloud is longer than the disc dynamical time). The gas clouds
coagulate slowly, and at the same time they are sheared apart by the disc
differential rotation. This slow coagulation process can produce clouds
characterized by power-law mass distributions where small clouds dominate in
number \citet{1983ApJ...271..604K} \footnote{We note, however, that this can
also be produced by theory assuming turbulence fragmentation 
\citep[e.g.][]{2013MNRAS.430.1653H}. However they assumed an barotropic
equation of state, and did not treat the multi-phase ISM.}.
In this regime, one expects to see only a small fraction of clouds undergoing
cloud-cloud collisions.
In fact, such collision candidates have been reported in the literate
\citep{2011A&A...528A..50D,2012ApJ...746...25N,2015ApJ...806....7T,2017arXiv170104669F,2017ApJ...835L..14G}.
 Such collisions can also produce rotating clouds, like those already observed
 in \citet{2017A&A...598A..96L,2017A&A...597A..70L}. But we
 should note that in the majority of the cases one should expect to observe a  smaller cloud nudging a larger cloud
\citep{2015MNRAS.446.3608D}.
Because of shear, one should also see long gas filaments that are parallel to the disc mid-plane.
These filaments are also seen
\citep{2013A&A...559A..34L,2014ApJ...797...53G,2014A&A...568A..73R,2015MNRAS.450.4043W,2015ApJ...815...23Z,2016ApJS..226....9W,2016A&A...591A...5L,2016A&A...590A.131A},
and are predicted by theories
\citep{2001MNRAS.327..663P,2014MNRAS.441.1628S,2015MNRAS.447.3390D}.

 At some high-density regions
such as the spiral arms, $\Sigma_{\rm   disc}$ starts to approach $1/3
\Sigma_{\rm cloud}$ and the velocity dispersion is also locally enhanced due to
the spiral density wave. As a result, cloud-cloud collisions become frequent.
 Molecular clouds start to  collide frequently within a relatively
short time. The cloud
collision cooling can  reduce the overall velocity
dispersion of molecular gas in the discs which leads to dynamical cooling, and
this dynamical cooling produces the complexes. However, in this particular case,
for cloud-cloud collisions to be frequent enough,
one needs processes such as the spiral density wave to enhance the density
beforehand.
As a result, these structures are formed along the spiral arms. It is believed that wiggle-like
regular-spaced structures seem on the arms of the grand-design spiral galaxy
M51 are produced in this way \citep{2008MNRAS.391..844D} \footnote{We note that
there are alternative explanation to these wiggles, where they come from the wiggle
(Kelvin-Helmholtz) instability
\citep{2004MNRAS.349..270W,2006MNRAS.367..873D,2006ApJ...647..997S,2006ApJ...646..213K}.
It seems unclear how Kelvin-Helmholtz instability can develop in such a multi-phased ISM where the equation of states are poorly defined. Therefore,
we prefer the explanation provided \citet{2008MNRAS.391..844D} where these spurs
 are produces by cloud collisions. }.
 Molecular complexes in the Milky Way such as W43
and W51 might also belong to this category. In this  regime, cloud collision
cooling has some influences on the disc dynamics. But these influences are
restricted to the close vicinity of the spiral arms where the densities are
enhanced.

For a massive, gas-rich disc, the cloud collision time is much shorter
than the disc dynamical time ($D>> 1/3$, $\Sigma_{\rm disc} >> 1/3 \Sigma_{\rm
cloud}$ \footnote{Where, if one could resolve these clouds in a galaxy, one should see the clouds overlap along the line
of sights.}, typically the cloud collision time is shorter than the disc
dynamical time by a factor of 3--10) at almost everywhere in the disc.
Cloud collision cooling rapidly removes kinetic energy from the disc and allows
the high-density regions (typically of $\sim$ kpc size) to contract from the
large scale. The contraction retrospectively enhances the cloud collision rate.
This positive feedback enables the masses of high-density regions to grow
rapidly, and it leads to the formation of clumps. The clump formation 
is thus an non-linear process, where cloud collision cooling starts to
back-react on the disc dynamics on the large scale and enables these large
perturbations to grow.
In this cooling-efficient regime, one should be able to observe clumps whose
sizes are comparable to the Toomre length \citep[][$l_{\rm Toomre}\approx 2 \uppi
G \Sigma_{\rm disc}/ \Omega_{\rm disc}^2$, which characterises the size of the
large-scale perturbations in the discs.
Perturbations larger than the Toomre length are suppressed due to
shear]{1964ApJ...139.1217T}, as has been reported by a recent observation
\citep{2017arXiv170300458F}.  We note that since the clumps are already the
end results of the cloud collision cooling process, their surface densities
should be much larger than $\Sigma_{\rm cloud}$, where we also expect the
clumps to contain a significant amount of overlapping sub-structures, as has
been seen by \citet{2017MNRAS.464..491F}.

\section{Conclusions}

We study the evolution of galactic discs, and aim to understand the
formation mechanism of the kpc-sized giant clumps commonly observed in gas-rich
discs. { We argue that for these discs, self-regulation leads to a Toomre
$Q$ parameter that is close to unity, and on top of this, efficient
dynamical cooling is a necessary condition for clump formation.

We approximate a galactic disc ISM as a medium that contains an ensemble of
colliding clouds, and study its evolution. We propose that
the major dynamical cooling channel of such a galactic disc is the
coagulation (collision) between molecular clouds. When the cloud-cloud
collision rate is high enough, this dynamical cooling (called cloud collision
cooling, see also \citet{1989ApJ...344..306E}) can back-react on the
disc, which enables it to contract on the large scale.
In spiral galaxies this produces molecular complexes seen on spiral arms, and in gas-rich discs this produces kpc-sized
molecular clumps.  }

%  On the disc scale, this
% constitutes an efficient way to remove kinetic energy.
%  Cloud collision cooling is the major mechanism
% that leads to clump formation.   Compared to cloud collision cooling, radiative
% cooling is not important for the global stability of the discs, as it merely removes the thermal energy
% and drives phase transitions, but cannot directly remove the kinetic energy of
% the disc.
% 

The effectiveness of cloud collision cooling is characterized by the $D$
(dissipation) parameter:
\begin{equation}\nonumber
D = \frac{t_{\rm cool}^{-1}}{t_{\rm ff}^{-1}}=\frac{\rho_{\rm
disc}^{1/2}\sigma_{\rm v, disc}}{G^{1/2} \Sigma_{\rm cloud}}\approx \frac{\Sigma_{\rm
disc}}{\Sigma_{\rm cloud}} \approx \frac{1}{12 Z }
\frac{\Sigma_{\rm disc}}{M_{\odot}\;\rm pc^{-2}}\;,
\end{equation}
where $\Sigma_{\rm disc}$ is the mean surface density of the galactic disc, and 
 $\Sigma_{\rm cloud}$ is the typical surface density of a molecular cloud in the
 disc, and $Z$ is the metallicity. The $D$ parameter is a fundamental parameter
 that determines the disc evolution: When  $D < 1/3$ ($t_{\rm cool} > 3
 \Omega_{\rm kep}^{-1}$, and roughly $\Sigma_{\rm  disc} < 1/3
\Sigma_{\rm cloud}$), the disc should enter a state where
 shear stretches gas into long, filamentary structures.
When  $D > 1/3$ ($t_{\rm cool}  < 3 \Omega_{\rm kep}^{-1}$, roughly $\Sigma_{\rm  disc} > 1/3
\Sigma_{\rm cloud}$), 
molecular clouds in a disc form a system that is highly dissipative where
energy is dissipated due to cloud collision cooling (cloud-cloud collisions),
and the disc should fragment into clumps/complexes.
 Since cloud-cloud collisions have played a vital role, the
 clumps/complexes formed in this way should contain a significant amount of
 overlapping sub-structures reminiscent of the colliding clouds. In our
 picture, the collisions have played different roles on different scales: on
 large scales, cloud collision cooling can back-react on the disc and enables
 the formation of giant clumps commonly observed in gas-rich discs, and on smaller scales, cloud-cloud collisions drive molecular cloud turbulence. 
In a more general sense, we have made an effort to understand how the
multi-phase ISM dissipates its kinetic energy. The current analysis is
carried out in a simplified model, and the dissipation rate estimates are accurate only
in the order of magnitude sense.  Nevertheless, when the
gas is separated into different phases, we expect system to be
almost-dissipationless when $D << 1/3$ and dissipative when $D >> 1/3$, and we
expect the transition to occur when the surface density of the galactic disc starts to
exceed the mean surface density of the molecular clouds by much. Understanding this transition
is of crucial importance for understanding the interplay between the ISM and the
disc dynamics, and this issue deserves further studies.

Theoretically, one can derive the
cloud collision cooling  rate using local, shearingbox-like simulations. This
would be a crucial test to our model. Once this is verified to a better accuracy, the
cooling time estimates can be incorporated into simulations of lower resolutions as subgrid models, and into
analytical models to study the fragmentation in different regimes. 
Such approaches have been previously made \citet{2007MNRAS.376.1588B}, and
one should make parameter studies and extend it to simulate formation of the
clumps. 

 Observationally, one
can  test our scenario by constraining the properties of
the molecular clouds and study their interactions in different galactic
environments.
The properties of the clouds can be constrained by high-resolution CO
observations. The global energy budged can be verified by comparing the cloud
velocity dispersions with the predictions made in Sec. \ref{sec:balance}. The
actual cloud collision process can be traced using transitions of shock-tracing
molecules such as SiO  and methanol (see also the prediction from a recent
chemical study by \citet{2017arXiv170607006B}).
Systematic studies of cloud evolution along these directions will help to
constrain the dynamical cooling and understand disc evolution.  \\

% Our model synthesises the two seemingly-contradicting views of clump formation.
% It was believed that either  clumps can from through the growth of perturbations
% on the scale of the disc scaleheight $H_{\rm disc}$ from a top-down fashion
% \citep{2009Natur.457..451D,2005MNRAS.363....2K}, or that clumps would form
% through the mergers of
% smaller structures
% \citep{2016ApJ...819L...2B,2015MNRAS.448.1007B}. In fact, there is no real
% contradiction: It is precisely the
% coagulation cooling occurs on small scale that enables the clumps to form
% through the growth of large-scale perturbations in the disc.\\

{\it Acknowledgements}
 Guang-Xing Li thanks Xun Shi for discussions and
for comments on the paper, and thanks Manuel Behrendt and Andi Burket for
introducing him the problem and for sharing their views.  Alexei Kritsuk is
acknowledged  for a discussion at MIAPP (Munich) where he introduced the
 \citet{1983Ap&SS..89..177M} paper, and Masataka Fukugita is acknowledged for a
 discussion on H$_{\rm I}$/H$_2$ transition. Jens Kauffman and Pavel Kroupa are
 acknowledge for critical comments. Bruce Elmegreen is acknowledged for an
 in-depth discussion on the multi-phase ISM. Finally, the referee must be
 thanked the a careful reading of the paper, and for the pertinent comments.
This paper is supported by the Deutsche Forschungsgemeinschaft (DFG) priority
program 1573 ISM-SPP.
\appendix
\section{Disc scaleheight corrections}\label{sec:h}
To convert $D=\rho_{\rm
disc}^{1/2}\sigma_{\rm v, disc}/G^{1/2} \Sigma_{\rm cloud} $ to the approximate
expression $D \approx \Sigma_{\rm disc } / \Sigma_{\rm cloud}$, one needs to
estimate the thickness of the molecular disc  (Eq.
\ref{eq:H})
\begin{equation}\label{eq:hh}
H_{\rm disc} = \frac{f_{\rm H} \sigma_{\rm v,  disc}}{\Omega_{\rm kep}}\;. 
\end{equation}
The derivation and error analysis of Eq. \ref{eq:hh} is present below.

We have assumed that the
scaleheight of the disc is determined by the balance between pressure in the disc, which is of order
\begin{equation} p_{\rm disc} \approx \rho_{\rm disc} \sigma_{\rm v,
disc}^2\approx (\Sigma_{\rm disc} / H_{\rm disc})\times   \sigma_{\rm v,
disc}^2\;,
\end{equation}
 and the vertical compression
due to the gravity from the dark matter 
\begin{equation}
p_{\rm gravity} \approx
a_{\rm z} \; \Sigma_{\rm disc}\approx
\Omega_{\rm kep}^2  r  \; (H_{\rm disc}/ r)\; \Sigma_{\rm disc} \;,
\end{equation}
where $a_z$
is the gravitational acceleration along the vertical direction. Here, the have
used the galaxy rotation velocity to estimate the acceleration (where $\rm r$ is
the radial direction and $z$ is the height).
$a_{\rm r} = \Omega_{\rm kep}^2  r  $, and $a_{\rm z} \approx H_{\rm disc}/ r\
\times a_{\rm r}$.

One can estimate the scaleheight of the disc using using $p_{\rm disc}
\approx p_{\rm gravity}$, which gives
\begin{equation}\label{eq:hhh}
H_{\rm disc} = \frac{\sigma_{\rm v,  disc}}{\Omega_{\rm kep}}\;,
\end{equation}
where one has neglected the self-gravity of the molecular
gas and the gravitational force from the stars.

We can prove that these effects are not important: the pressure due to
self-gravity is $p_{\rm self-gravity} \approx G \Sigma_{\rm disc}^2$, and one can prove that 
\begin{equation}
\frac{p_{\rm self-gravity}}{p_{\rm gravity}} \approx \frac{1}{Q_{\rm gas}}\;,
\end{equation}
where $Q_{\rm gas}$ is the defined as 
\begin{equation}
Q_{\rm gas} = \frac{\sigma_{\rm v, gas} \Omega_{\rm kep}}{\uppi G \Sigma_{\rm
gas}}\;.
\end{equation}
It is easy to see that self-gravity is not important when $Q_{\rm gas} >1$, and
is only comparable to the gravitational force from the dark matter halo when
$Q_{\rm gas} =1$. When $Q_{\rm gas} >>1$, Eq. \ref{eq:hh} provides an extremely
good estimate to the disc thickness, and when $Q_{\rm gas}\approx 1$,  Eq.
\ref{eq:hh} overestimates the disc scaleheight by $\sqrt{2}-1\approx 0.4$. In
either cases the error is not significant. The galaxies that we are studying 
have $Q_{\rm gas} \gtrsim 1$ \citep{2013MNRAS.433.1389R,2017arXiv170102138R},
and as a result Eq. \ref{eq:hh} is always a good estimate to the disc
thickness. In additional to these, the gravitational force from the stars can
further reduce the disc thickness, but after going through a similar analysis,
one can prove that the effect is also negligible for our purpose.
\bibliography{paper}
%%%%%%%%%%%%%%%%%%%%%%%%%%%%%%%%%%%%%%%%%%%%%%%%%%

% Don't change these lines
\bsp	% typesetting comment
\label{lastpage}
\end{document}